\documentclass[12pt]{article}
\usepackage{graphicx}
\usepackage{amsmath}
\usepackage{amssymb}
\usepackage{caption2}
\begin{document}
\bibliographystyle{prsty}
\begin{center}
{\large {\bf \sc{  Analysis of  $\Omega_c^*(css)$ and $\Omega_b^*(bss)$  with QCD sum rules }}} \\[2mm]
Zhi-Gang Wang \footnote{E-mail,wangzgyiti@yahoo.com.cn;wangzg@yahoo.cn.  }     \\
 Department of Physics, North China Electric Power University,
Baoding 071003, P. R. China

\end{center}

\begin{abstract}
In this article, we calculate the masses and residues
  of the heavy  baryons $\Omega_c^*(css)$ and $\Omega_b^*(bss)$
  with spin-parity $\frac{3}{2}^+$ with the QCD sum rules.
  The  numerical values are  compatible  with experimental data and other theoretical
  estimations.
\end{abstract}

 PACS number: 14.20.Lq, 14.20.Mr

Key words: $\Omega_c^*(css)$,  $\Omega_b^*(bss)$, QCD sum rules

\section{Introduction}

Several new excited charmed baryon states have been observed by
BaBar, Belle and CLEO Collaborations, such as $\Lambda_c(2765)^+$,
$\Lambda_c^+(2880)$, $\Lambda_c^+(2940)$, $\Sigma_c^+(2800)$,
$\Xi_c^+(2980)$, $\Xi_c^+(3077)$,
  $\Xi_c^0(2980)$ , $\Xi_c^0(3077)$ \cite{NewBaryon,ShortRV}. The charmed
baryons  provide a rich source of states, including  possible
candidates for the orbital excitations. They serve as an excellent
ground for testing  predictions of the constituent  quark models and
heavy quark symmetry \cite{ReviewH}. The charmed and bottomed
baryons, which contain a heavy quark and two light quarks,  provides
an ideal tool for studying  dynamics of the light quarks in the
presence  of a heavy quark. The $u$, $d$ and $s$ quarks form an
$SU(3)$ flavor triplet, ${\bf 3}\times {\bf 3}={\bf \bar 3}+{\bf
6}$, two light quarks can form diquarks with a symmetric sextet  and
an antisymmetric antitriplet. For the $S$-wave baryons, the sextet
contains both spin-$\frac{1}{2}$ and spin-$\frac{3}{2}$ states,
while the antitriplet contains only spin-$\frac{1}{2}$ states. By
now, the ${1\over 2}^+$ antitriplet states ($\Lambda_c^+$,
$\Xi_c^+,\Xi_c^0)$,  and the ${1\over 2}^+$ and ${3\over 2}^+$
sextet states  ($\Omega_c,\Sigma_c,\Xi'_c$) and
($\Omega_c^*,\Sigma_c^*,\Xi'^*_c$) have been established.

The baryon $\Omega_{c}^{*}$, a $css$ candidate for the $\frac{3}{2}$
partner of the strange baryon $\Omega(sss)$, was observed by  BaBar
collaboration in
 the radiative decay $\Omega_{c}^{*}\rightarrow\Omega_c\gamma$ \cite{OmegaC}. The
${1\over 2}^+$ baryon $\Omega_c(css)$ was reconstructed in decays to
the final states $\Omega^-\pi^+$, $\Omega^-\pi^+\pi^0$,
$\Omega^-\pi^+\pi^-\pi^+$ and $\Xi^- K^-\pi^+\pi^+$. It lies about
$70.8 \pm 1.0 \pm 1.1 \rm{MeV}$ above the $\Omega_c$, and it is the
last singly-charmed baryon  with zero orbital momentum  observed
experimentally \cite{Rosner95}.

In this article, we  calculate the mass and residue of the
 $\Omega_c^*$ (and $\Omega_b^*$ as byproduct, the $\Omega_b^*$ has not been observed experimentally yet)
 with the QCD sum rules
\cite{SVZ79,Narison89}. In the QCD sum rules, operator product
expansion is used to expand the time-ordered currents into a series
of quark and gluon condensates  which parameterize the long distance
properties of  the QCD vacuum. Based on current-hadron duality, we
can obtain copious information about the hadronic parameters at the
phenomenological side.

The article is arranged as follows:  we derive the QCD sum rules for
the masses and residues of  the $\Omega_c^*$ and $\Omega_b^*$ in
section 2; in section 3, numerical results and discussions; section
4 is reserved for conclusion.

\section{QCD sum rules for  the $\Omega_c^*$ and $\Omega_b^*$}
In the following, we write down  the two-point correlation functions
$\Pi^a_{\mu\nu}(p^2)$ in the QCD sum rules approach,
\begin{eqnarray}
\Pi^a_{\mu\nu}(p^2)&=&i\int d^4x e^{ip \cdot x} \langle
0|T\left\{J^{a}_\mu(x)\bar{J}^{a}_\nu(0)\right\}|0\rangle \, ,  \\
J^{a}_\mu(x)&=&\epsilon_{ijk}s^T_i(x)C\gamma_\mu s_j(x) Q^a_k(x) \, ,\\
\lambda_{a}N_\mu(p,s)  &=& \langle
0|J^{a}_\mu(0)|\Omega_a^*(p,s)\rangle\, ,
\end{eqnarray}
where the upper index $a$ represents the $c$ and $b$ quarks
respectively; the $N_\mu(p,s)$ and $\lambda_{a}$ stand for the
Rarita-Schwinger spin vector and residue of the baryon $\Omega_a^*$,
respectively. $i$, $j$ and $k$ are color indexes, $C$ is charge
conjunction matrix,  and $\mu$ and $\nu$ are Lorentz indexes.

The correlation functions  $\Pi^a_{\mu\nu}(p)$ can be decomposed as
follows:
\begin{eqnarray}
\Pi^a_{\mu\nu}(p)&=&-g_{\mu\nu} \left\{\!\not\!{p}\Pi^a_1(p^2)
+\Pi_2^a(p^2) \right\}+\cdots,
\end{eqnarray}
due to  Lorentz covariance.  The first structure
$g_{\mu\nu}\!\not\!{p}$ has an odd number of $\gamma$-matrices  and
conserves chirality, the second  structure $g_{\mu\nu}$ has an even
number of $\gamma$-matrices and violate chirality. In the original
QCD sum rules analysis of the nucleon masses and magnetic moments
\cite{Ioffe81}, the interval of dimensions (of the condensates) for
the odd structure is larger than the interval of dimensions for the
even structure, one may expect a better accuracy of  results
obtained from the sum rules with  the odd structure.

In this article, we choose the two tensor structures  to study the
masses and residues of the  heavy baryons $\Omega_c^*$ and
$\Omega_b^*$, as the masses of the heavy quarks break the chiral
symmetry explicitly.

 According to   basic assumption of current-hadron duality in
the QCD sum rules approach \cite{SVZ79}, we insert  a complete
series of intermediate states satisfying  unitarity   principle with
the same quantum numbers as the current operator $J^{a}_\mu(x)$
 into the correlation functions in
Eq.(1)  to obtain the hadronic representation. After isolating the
pole terms  of the lowest states $\Omega_a^*$, we obtain the
following result:
\begin{eqnarray}
\Pi^a_{\mu\nu}(p^2)&=&-g_{\mu\nu}\lambda_{a}^2\frac{M_{\Omega_a^*}+\!\not\!{p}}{M_{\Omega_a^*}^2-p^2}
+\cdots \, \, ,
\end{eqnarray}
where we have used the relation to sum over the Rarita-Schwinger
spin vector,
\begin{eqnarray}
\sum_s N_\mu(p,s) \bar{N}_\nu(p,s)
=-(\!\not\!{p}+M_{\Omega_a^*})\left\{ g_{\mu\nu}-\frac{\gamma_\mu
\gamma_\nu}{3}-\frac{2p_\mu p_\nu}{3M_{\Omega_a^*}^2}+\frac{p_\mu
\gamma_\nu-p_\nu \gamma_\mu}{3M_{\Omega_a^*}} \right\} \, .
\end{eqnarray}

In the following, we briefly outline  operator product expansion for
the correlation functions $\Pi^a_{\mu\nu }(p)$  in perturbative QCD
theory. The calculations are performed at   large space-like
momentum region $p^2\ll 0$, which corresponds to small distance
$x\approx 0$ required by   validity of   operator product expansion.
We write down the "full" propagators $S_{ij}(x)$ and $S_Q^{ij}(x)$
of a massive quark in the presence of the vacuum condensates firstly
\cite{SVZ79}\footnote{One can consult the last article of
Ref.\cite{SVZ79} for technical details in deriving the full
propagator.},
\begin{eqnarray}
S_{ij}(x)&=& \frac{i\delta_{ij}\!\not\!{x}}{ 2\pi^2x^4}
-\frac{\delta_{ij}m_s}{4\pi^2x^2}-\frac{\delta_{ij}}{12}\langle
\bar{s}s\rangle +\frac{i\delta_{ij}}{48}m_s
\langle\bar{s}s\rangle\!\not\!{x}-\frac{\delta_{ij}x^2}{192}\langle \bar{s}g_s\sigma Gs\rangle\nonumber\\
&& +\frac{i\delta_{ij}x^2}{1152}m_s\langle \bar{s}g_s\sigma
 Gs\rangle \!\not\!{x}-\frac{i}{32\pi^2x^2} G^{ij}_{\mu\nu} (\!\not\!{x}
\sigma^{\mu\nu}+\sigma^{\mu\nu} \!\not\!{x})  +\cdots \, ,\nonumber\\
S_Q^{ij}(x)&=&\frac{i}{(2\pi)^4}\int d^4k e^{-ik \cdot x} \left\{
\frac{\delta_{ij}}{\!\not\!{k}-m_Q}
-\frac{g_sG^{\alpha\beta}_{ij}}{4}\frac{\sigma_{\alpha\beta}(\!\not\!{k}+m_Q)+(\!\not\!{k}+m_Q)
\sigma_{\alpha\beta}}{(k^2-m_Q^2)^2}\right.\nonumber\\
&&\left.+\frac{\pi^2}{3} \langle \frac{\alpha_sGG}{\pi}\rangle
\delta_{ij}m_Q \frac{k^2+m_Q\!\not\!{k}}{(k^2-m_Q^2)^4}
+\cdots\right\} \, ,
\end{eqnarray}
where $\langle \bar{s}g_s\sigma Gs\rangle=\langle
\bar{s}g_s\sigma_{\alpha\beta} G^{\alpha\beta}s\rangle$  and
$\langle \frac{\alpha_sGG}{\pi}\rangle=\langle
\frac{\alpha_sG_{\alpha\beta}G^{\alpha\beta}}{\pi}\rangle$, then
contract the quark fields in the correlation functions
$\Pi^a_{\mu\nu}(p)$ with Wick theorem, and obtain the result:
\begin{eqnarray}
\Pi^a_{\mu\nu}(p)&=&2i \epsilon_{ijk}\epsilon_{i'j'k'} \int d^4x \,
e^{i p \cdot x} Tr\left\{ \gamma_\mu S_{ii'}(x)\gamma_\nu C
S^T_{jj'}(x)C\right\}S_Q^{kk'}(x)\, .
\end{eqnarray}
Substitute the full $s$, $c$ and $b$ quark propagators into above
correlation functions and complete  the integral in  coordinate
space, then integrate over the variable $k$, we can obtain the
correlation functions $\Pi^a_i(p^2)$ at the level of quark-gluon
degree of freedom:

\begin{eqnarray}
\Pi_1^a(p^2)&=& -\frac{1}{64\pi^4} \int_0^1 dx x(1-x)^2(x+2)\left(
\widetilde{m}_a^2-p^2\right)^2 \log\left(
\widetilde{m}_a^2-p^2\right)
\nonumber \\
&&-\frac{m_s\langle \bar{s}s\rangle}{4\pi^2}\int_0^1 dx
x(x-2)\log\left( \widetilde{m}_a^2-p^2\right) \nonumber \\
&&+\frac{1}{192\pi^2}\langle \frac{\alpha_sGG}{\pi}\rangle \int_0^1
dx  x(x-2) \log\left(\widetilde{m}_a^2-p^2\right)\nonumber\\
&&+\frac{m_s\langle \bar{s}g_s\sigma Gs\rangle}{24\pi^2} \int_0^1 dx
\frac{x}{\widetilde{m}_a^2-p^2}-\frac{m_a^2}{576\pi^2}\langle
\frac{\alpha_sGG}{\pi}\rangle \int_0^1 dx
\frac{(1-x)^2(x+2)}{x^2(\widetilde{m}_a^2-p^2)} \nonumber\\
&&+\frac{\langle \bar{s}s\rangle^2}{3}
\frac{1}{m_a^2-p^2}+\frac{m_s\langle \bar{s}g_s\sigma
Gs\rangle}{12\pi^2} \frac{1}{m_a^2-p^2}+\cdots \, \, , \\
\Pi_2^a(p^2)&=& -\frac{m_a}{64\pi^4} \int_0^1 dx (1-x)^2(x+2)\left(
\widetilde{m}_a^2-p^2\right)^2 \log\left(
\widetilde{m}_a^2-p^2\right)
\nonumber \\
&&-\frac{m_am_s\langle \bar{s}s\rangle}{4\pi^2}\int_0^1 dx
(x-2)\log\left( \widetilde{m}_a^2-p^2\right) \nonumber \\
&&-\frac{m_a}{576\pi^2}\langle \frac{\alpha_sGG}{\pi}\rangle
\int_0^1
dx  (\frac{4}{x^2}-\frac{9}{x}-3x^2+2x+9) \log\left(\widetilde{m}_a^2-p^2\right)\nonumber\\
&&+\frac{m_a}{576\pi^2}\langle \frac{\alpha_sGG}{\pi}\rangle
\int_0^1
dx \frac{x^4-x^3-3x^2+5x-2}{x(1-x)} \frac{\widetilde{m}_a^2 }{\widetilde{m}_a^2-p^2}\nonumber\\
&&+\frac{m_a m_s\langle \bar{s}g_s\sigma Gs\rangle}{24\pi^2}
\int_0^1 dx \frac{1}{\widetilde{m}_a^2 -p^2}+\frac{m_a\langle
\bar{s}s\rangle^2}{3} \frac{1}{m_a^2-p^2}
\nonumber\\
&&+\frac{m_am_s\langle \bar{s}g_s\sigma Gs\rangle}{12\pi^2}
\frac{1}{m_a^2-p^2}+\cdots \, \, ,
\end{eqnarray}
where $\widetilde{m}_a^2=\frac{m_a^2}{x}$.

  We carry out  operator
product expansion to the vacuum condensates adding up to
dimension-6. In calculation, we
 take  assumption of vacuum saturation for  high
dimension vacuum condensates, they  are always
 factorized to lower condensates with vacuum saturation in the QCD sum rules,
  factorization works well in  large $N_c$ limit.
In this article, we take into account the contributions from the
quark condensate $\langle \bar{s}s \rangle$,  mixed condensate
$\langle \bar{s}g_s \sigma  G{s} \rangle $, gluon condensate
$\langle \frac{\alpha_s GG}{\pi}\rangle$, and neglect the
contributions  from other high dimension condensates, which are
suppressed by large denominators and would not play significant
roles.

Once  analytical results are obtained,
  then we can take  current-hadron duality  below the threshold
$s^0_{a}$ and perform  Borel transformation with respect to the
variable $P^2=-p^2$, finally we obtain  the following sum rules:

\begin{eqnarray}
\lambda^2_{a}\exp\left\{-\frac{M_{\Omega_a^*}^2}{M^2}\right\}&=&
\frac{1}{64\pi^4} \int_{th}^{s^0_{a}}ds \int_{\Delta^a}^1 dx
x(1-x)^2(x+2)\left( \widetilde{m}_a^2-s\right)^2
\exp\left\{-\frac{s}{M^2}\right\}
\nonumber \\
&&+\frac{m_s\langle
\bar{s}s\rangle}{4\pi^2}\int_{th}^{s^0_{a}}ds\int_{\Delta^a}^1 dx
x(x-2)\exp\left\{-\frac{s}{M^2}\right\} \nonumber \\
&&-\frac{1}{192\pi^2}\langle \frac{\alpha_sGG}{\pi}\rangle
\int_{th}^{s^0_{a}}ds\int_{\Delta^a}^1
dx  x(x-2) \exp\left\{-\frac{s}{M^2}\right\}\nonumber\\
&&+\frac{m_s\langle \bar{s}g_s\sigma Gs\rangle}{24\pi^2} \int_0^1 dx
x\exp\left\{-\frac{\widetilde{m}_a^2}{M^2}\right\}\nonumber\\
&&-\frac{m_a^2}{576\pi^2}\langle \frac{\alpha_sGG}{\pi}\rangle
\int_0^1 dx
\frac{(1-x)^2(2+x)}{x^2}\exp\left\{-\frac{\widetilde{m}_a^2}{M^2}\right\} \nonumber\\
&&+\frac{\langle \bar{s}s\rangle^2}{3}
\exp\left\{-\frac{m_a^2}{M^2}\right\}+\frac{m_s\langle
\bar{s}g_s\sigma Gs\rangle}{12\pi^2}
\exp\left\{-\frac{m_a^2}{M^2}\right\} \, \, ,
\end{eqnarray}

\begin{eqnarray}
&&M_{\Omega^*_a}\lambda^2_{a}\exp\left\{-\frac{M_{\Omega_a^*}^2}{M^2}\right\}
\nonumber\\
 &=& \frac{m_a}{64\pi^4} \int_{th}^{s_a^0}
ds\int_{\Delta^a}^1 dx (1-x)^2(x+2)\left(
\widetilde{m}_a^2-s\right)^2\exp\left\{-\frac{s}{M^2}\right\}
\nonumber \\
&&+\frac{m_am_s\langle \bar{s}s\rangle}{4\pi^2}\int_{th}^{s_a^0} ds
\int_{\Delta^a}^1 dx (x-2)\exp\left\{-\frac{s}{M^2}\right\} \nonumber \\
&&+\frac{m_a}{576\pi^2}\langle \frac{\alpha_sGG}{\pi}\rangle
\int_{th}^{s_a^0} ds \int_{\Delta^a}^1
dx  (\frac{4}{x^2}-\frac{9}{x}-3x^2+2x+9) \exp\left\{-\frac{s}{M^2}\right\} \nonumber\\
&&+\frac{m_a}{576\pi^2}\langle \frac{\alpha_sGG}{\pi}\rangle
\int_0^1
dx \frac{x^4-x^3-3x^2+5x-2}{x(1-x)} \widetilde{m}_a^2 \exp\left\{-\frac{\widetilde{m}_a^2}{M^2}\right\}\nonumber\\
&&+\frac{m_a m_s\langle \bar{s}g_s\sigma Gs\rangle}{24\pi^2}
\int_0^1 dx
\exp\left\{-\frac{\widetilde{m}_a^2}{M^2}\right\}+\frac{m_a\langle
\bar{s}s\rangle^2}{3} \exp\left\{-\frac{m^2_a}{M^2}\right\}
\nonumber\\
&&+\frac{m_am_s\langle \bar{s}g_s\sigma Gs\rangle}{12\pi^2}
\exp\left\{-\frac{m^2_a}{M^2}\right\} \, \, ,
\end{eqnarray}
where $th=(m_a+2m_s)^2$ and $\Delta^a=\frac{m_a^2}{s}$.

Differentiate the above sum rules with respect to the variable
$\frac{1}{M^2}$, then eliminate the quantity $\lambda_{\Omega_a^*}$,
we obtain two QCD sum rules for the masses $M_{\Omega_a^*}$:

\begin{eqnarray}
M_{\Omega_a^*}^2&=& \left\{\frac{1}{64\pi^4} \int_{th}^{s^0_{a}}ds
\int_{\Delta^a}^1 dx x(1-x)^2(x+2)\left(
\widetilde{m}_a^2-s\right)^2 s \exp\left\{-\frac{s}{M^2}\right\}
\right.
\nonumber \\
&&+\frac{m_s\langle
\bar{s}s\rangle}{4\pi^2}\int_{th}^{s^0_{a}}ds\int_{\Delta^a}^1 dx
x(x-2)s \exp\left\{-\frac{s}{M^2}\right\} \nonumber \\
&&-\frac{1}{192\pi^2}\langle \frac{\alpha_sGG}{\pi}\rangle
\int_{th}^{s^0_{a}}ds\int_{\Delta^a}^1
dx  x(x-2)s \exp\left\{-\frac{s}{M^2}\right\}\nonumber\\
&&+\frac{m_sm_a^2\langle \bar{s}g_s\sigma Gs\rangle}{24\pi^2}
\int_0^1 dx
 \exp\left\{-\frac{\widetilde{m}_a^2}{M^2}\right\}\nonumber\\
&&-\frac{m_a^4}{576\pi^2}\langle \frac{\alpha_sGG}{\pi}\rangle
\int_0^1 dx
\frac{(1-x)^2(2+x)}{x^3}\exp\left\{-\frac{\widetilde{m}_a^2}{M^2}\right\} \nonumber\\
&&\left.+\frac{m_a^2\langle \bar{s}s\rangle^2}{3}
\exp\left\{-\frac{m_a^2}{M^2}\right\}+\frac{m_sm_a^2\langle
\bar{s}g_s\sigma Gs\rangle}{12\pi^2}
\exp\left\{-\frac{m_a^2}{M^2}\right\}\right\}/\nonumber\\
&&\left\{\frac{1}{64\pi^4} \int_{th}^{s^0_{a}}ds \int_{\Delta^a}^1
dx x(1-x)^2(x+2)\left( \widetilde{m}_a^2-s\right)^2
\exp\left\{-\frac{s}{M^2}\right\} \right.
\nonumber \\
&&+\frac{m_s\langle
\bar{s}s\rangle}{4\pi^2}\int_{th}^{s^0_{a}}ds\int_{\Delta^a}^1 dx
x(x-2)\exp\left\{-\frac{s}{M^2}\right\} \nonumber \\
&&-\frac{1}{192\pi^2}\langle \frac{\alpha_sGG}{\pi}\rangle
\int_{th}^{s^0_{a}}ds\int_{\Delta^a}^1
dx  x(x-2) \exp\left\{-\frac{s}{M^2}\right\}\nonumber\\
&&+\frac{m_s\langle \bar{s}g_s\sigma Gs\rangle}{24\pi^2} \int_0^1 dx
x\exp\left\{-\frac{\widetilde{m}_a^2}{M^2}\right\}\nonumber\\
&&-\frac{m_a^2}{576\pi^2}\langle \frac{\alpha_sGG}{\pi}\rangle
\int_0^1 dx
\frac{(1-x)^2(x+2)}{x^2}\exp\left\{-\frac{\widetilde{m}_a^2}{M^2}\right\} \nonumber\\
&&\left.+\frac{\langle \bar{s}s\rangle^2}{3}
\exp\left\{-\frac{m_a^2}{M^2}\right\}+\frac{m_s\langle
\bar{s}g_s\sigma Gs\rangle}{12\pi^2}
\exp\left\{-\frac{m_a^2}{M^2}\right\} \right\} \, \, ,
\end{eqnarray}
and
\begin{eqnarray}
M_{\Omega^*_a}^2
 &=& \left\{\frac{m_a}{64\pi^4} \int_{th}^{s_a^0}
ds\int_{\Delta^a}^1 dx (1-x)^2(x+2)\left(
\widetilde{m}_a^2-s\right)^2s\exp\left\{-\frac{s}{M^2}\right\}
\right.
\nonumber \\
&&+\frac{m_am_s\langle \bar{s}s\rangle}{4\pi^2}\int_{th}^{s_a^0} ds
\int_{\Delta^a}^1 dx (x-2)s\exp\left\{-\frac{s}{M^2}\right\} \nonumber \\
&&+\frac{m_a}{576\pi^2}\langle \frac{\alpha_sGG}{\pi}\rangle
\int_{th}^{s_a^0} ds \int_{\Delta^a}^1
dx  (\frac{4}{x^2}-\frac{9}{x}-3x^2+2x+9) s\exp\left\{-\frac{s}{M^2}\right\} \nonumber\\
&&+\frac{m_a}{576\pi^2}\langle \frac{\alpha_sGG}{\pi}\rangle
\int_0^1
dx \frac{x^4-x^3-3x^2+5x-2}{x(1-x)} \widetilde{m}_a^4 \exp\left\{-\frac{\widetilde{m}_a^2}{M^2}\right\}\nonumber\\
&&+\frac{m_a^3 m_s\langle \bar{s}g_s\sigma Gs\rangle}{24\pi^2}
\int_0^1 dx \frac{1}{x}
\exp\left\{-\frac{\widetilde{m}_a^2}{M^2}\right\}+\frac{m_a^3\langle
\bar{s}s\rangle^2}{3} \exp\left\{-\frac{m^2_a}{M^2}\right\}
\nonumber\\
&&\left.+\frac{m_a^3m_s\langle \bar{s}g_s\sigma Gs\rangle}{12\pi^2}
\exp\left\{-\frac{m^2_a}{M^2}\right\}\right\} / \nonumber \\
&& \left\{\frac{m_a}{64\pi^4} \int_{th}^{s_a^0} ds\int_{\Delta^a}^1
dx (1-x)^2(x+2)\left(
\widetilde{m}_a^2-s\right)^2\exp\left\{-\frac{s}{M^2}\right\}
\right.
\nonumber \\
&&+\frac{m_am_s\langle \bar{s}s\rangle}{4\pi^2}\int_{th}^{s_a^0} ds
\int_{\Delta^a}^1 dx (x-2)\exp\left\{-\frac{s}{M^2}\right\} \nonumber \\
&&+\frac{m_a}{576\pi^2}\langle \frac{\alpha_sGG}{\pi}\rangle
\int_{th}^{s_a^0} ds \int_{\Delta^a}^1
dx  (\frac{4}{x^2}-\frac{9}{x}-3x^2+2x+9) \exp\left\{-\frac{s}{M^2}\right\} \nonumber\\
&&+\frac{m_a}{576\pi^2}\langle \frac{\alpha_sGG}{\pi}\rangle
\int_0^1
dx \frac{x^4-x^3-3x^2+5x-2}{x(1-x)} \widetilde{m}_a^2 \exp\left\{-\frac{\widetilde{m}_a^2}{M^2}\right\}\nonumber\\
&&+\frac{m_a m_s\langle \bar{s}g_s\sigma Gs\rangle}{24\pi^2}
\int_0^1 dx
\exp\left\{-\frac{\widetilde{m}_a^2}{M^2}\right\}+\frac{m_a\langle
\bar{s}s\rangle^2}{3} \exp\left\{-\frac{m^2_a}{M^2}\right\}
\nonumber\\
&&\left.+\frac{m_am_s\langle \bar{s}g_s\sigma Gs\rangle}{12\pi^2}
\exp\left\{-\frac{m^2_a}{M^2}\right\}\right\} \, \, .
\end{eqnarray}

\section{Numerical results and discussions}
The input parameters are taken to be the standard values $\langle
\bar{q}q \rangle=-(0.24\pm 0.01 \rm{GeV})^3$, $\langle \bar{s}s
\rangle=(0.8\pm 0.2 )\langle \bar{q}q \rangle$, $\langle
\bar{s}g_s\sigma Gs \rangle=m_0^2\langle \bar{s}s \rangle$,
$m_0^2=(0.8 \pm 0.2)\rm{GeV}^2$, $\langle \frac{\alpha_s
GG}{\pi}\rangle=(0.33\rm{GeV})^4 $, $m_s=(0.14\pm0.01)\rm{GeV}$,
$m_c=(1.4\pm0.1)\rm{GeV}$ and $m_b=(4.8\pm0.1)\rm{GeV}$
\cite{SVZ79,Narison89,Ioffe2005}.
 The contribution from the gluon condensate $\langle
\frac{\alpha_s GG}{\pi}\rangle $ is less than $4\%$, and the
uncertainty is neglected here.

For the octet baryons with $I(J^{P})=\frac{1}{2}({\frac{1}{2}}^+)$,
the mass of the proton (the ground state)  is $M_p=938\rm{MeV}$, and
the mass of the first radial excited state $N(1440)$ (the Roper
resonance) is $M_{1440}=(1420-1470)\rm{MeV}\approx 1440\rm{MeV}$
\cite{PDG}. For the decuplet  baryons with
$I(J^{P})=\frac{3}{2}({\frac{3}{2}}^+)$ , the mass of  the
$\Delta(1232)$ (the ground state) is
$M_{1232}=(1231-1233)\rm{MeV}\approx 1232\rm{MeV}$,  and the mass of
the first radial excited state $\Delta(1600)$ is
$M_{1600}=(1550-1700)\rm{MeV}\approx 1600\rm{MeV}$ \cite{PDG}. The
separation between the ground states and first radial excited states
is about $0.5\rm{GeV}$.  So in the QCD sum rules for the baryons
with the light quarks, the threshold parameters $s_0$ are always
chosen to be $\sqrt{s_0}=M_{gr}+0.5\rm{GeV}$
\cite{Ioffe81,Pasupathy85}, here $gr$ stands for the ground states.
The threshold parameters for the heavy baryons $\Omega_c^*$ and
$\Omega_b^*$ can be  chosen to be $s^0_{\Omega_c^*}=(2.8+0.5)^2
\rm{GeV}^2$ and $s^0_{\Omega_b^*} = (6.1+0.5)^2 \rm{GeV}^2$,
respectively. The mass of the bottomed baryon $\Omega_b^*$ with
spin-parity $\frac{3}{2}^+$ is about
$M_{\Omega_b^*}=(6.04-6.09)\rm{GeV}$, which is  predicted by the
quark models and lattice QCD \cite{QuarkMassB,LattMassB}.

\begin{table}
\begin{center}
\begin{tabular}{|c|c|c|}
\hline\hline & Eq.(11)& Eq.(12)\\ \hline
      $\mbox{perturbative term}$  &$+80\%$ &$+83\%$\\ \hline
      $ \langle \bar{s} s\rangle$& $+12\%$ &$+10\%$\\      \hline
     $ \langle \bar{s} g_s
\sigma G s\rangle$& $-4\%$ &$-2\%$\\     \hline
    $\langle\bar{s} s\rangle^2$&  $+12\%$ &$+7\%$\\ \hline
$\langle\frac{\alpha_s GG}{\pi}\rangle$&  $+1\%$ &$+2\%$\\ \hline
    \hline
\end{tabular}
\end{center}
\caption{ The contributions from different terms  in the sum rules
for the $\Omega_c^*$ with the central values of the input
parameters. }
\end{table}

\begin{table}
\begin{center}
\begin{tabular}{|c|c|c|}
\hline\hline & Eq.(11)& Eq.(12)\\ \hline
      $\mbox{perturbative term}$  &$+78\%$ &$+80\%$\\ \hline
      $ \langle \bar{s} s\rangle$& $+10\%$ &$+10\%$\\      \hline
     $ \langle \bar{s} g_s
\sigma G s\rangle$& $-4\%$ &$-3\%$\\     \hline
    $\langle\bar{s} s\rangle^2$&  $+15\%$ &$+12\%$\\ \hline
$\langle\frac{\alpha_s GG}{\pi}\rangle$&  $+1\%$ &$+1\%$\\ \hline
    \hline
\end{tabular}
\end{center}
\caption{ The contributions from different terms  in the sum rules
for the $\Omega_b^*$ with the central values of the input
parameters.}
\end{table}

In this article, the threshold parameters and Borel parameters are
taken as $s^0_{\Omega_c^*}=11.0\rm{GeV}^2$ and
$M^2=(2.5-3.5)\rm{GeV}^2$ for the charmed  baryon  $\Omega_c^*$, and
$s^0_{\Omega_b^*}=45.0\rm{GeV}^2$ and $M^2=(5.0-6.0)\rm{GeV}^2$ for
the bottomed baryon $\Omega_b^*$. The contributions from different
terms for the central values of the input parameters are presented
in Table.1 and Table.2, respectively. From the two tables, we can
expect convergence of the operator product expansion. In the two sum
rules in Eqs.(11-12), the contributions from the terms proportional
to the quark condensate $\langle \bar{s}s\rangle$ and mixed
condensate $\langle \bar{s}g_s \sigma G s\rangle$ are suppressed due
to the small mass $m_s$  comparing with the terms proportional to
the $\langle \bar{s}s\rangle^2$. Furthermore, from the 'full'
propagator of the $s$ quark, we can see that the mixed condensate
$\langle \bar{s}g_s \sigma G s\rangle$ is companied with additional
large denominators, its contribution is even smaller. In the
right-hand side of Eqs.(11-12), the terms proportional to the
$\langle \bar{s}s\rangle^2$ are suppressed by the exponents 
$\exp[-m^2_a/M^2]$, which is balanced by the factor
$\exp[-M_{\Omega_a^*}^2/M^2]$ in the left-hand side. Although the
masses of the $c$ quark and $\Omega_c^*$ baryon are much smaller
than the corresponding ones of the $b$ quark and $\Omega_b^*$
baryon, the Borel parameters $M^2$ are different, for the central
values of the Borel parameters $M^2$,  $\exp[
M_{\Omega_b^*}^2/M^2-m^2_b/M^2] > \exp
[M_{\Omega_c^*}^2/M^2-m^2_c/M^2]$. It is not unexpected, the
contributions from the $\langle \bar{s}s\rangle^2$ are larger in the
sum rules for the $\Omega_b^*$ baryon than the ones for the
$\Omega_c^*$ baryon.

If we approximate the phenomenological spectral density with the
perturbative term, the contribution from the pole term   is as large
as $(28-54) \%$ for the charmed baryon $\Omega_c^*$ and $(33-50) \%$
for the bottomed baryon $\Omega_b^*$. We can choose smaller Borel
parameter $M^2$ or larger threshold parameters $s^0_a$ to enhance
the contributions from the ground states. However, if we take larger
threshold parameter $s^0_a$, the contribution from the first radial
excited state maybe included in; on the other hand, for smaller
Borel  parameter $M^2$, the sum rules are not stable enough, the
uncertainty with variation of the Borel parameter is large.  In the
case of the multiquark states, the standard criterion of the lowest
pole dominance  cannot be satisfied, we have to resort to new
criterion to overcome the problem, for detailed discussions about
this subject, one can consult Ref.\cite{Wang06}.

\begin{figure}
 \centering
 \includegraphics[totalheight=6cm,width=7cm]{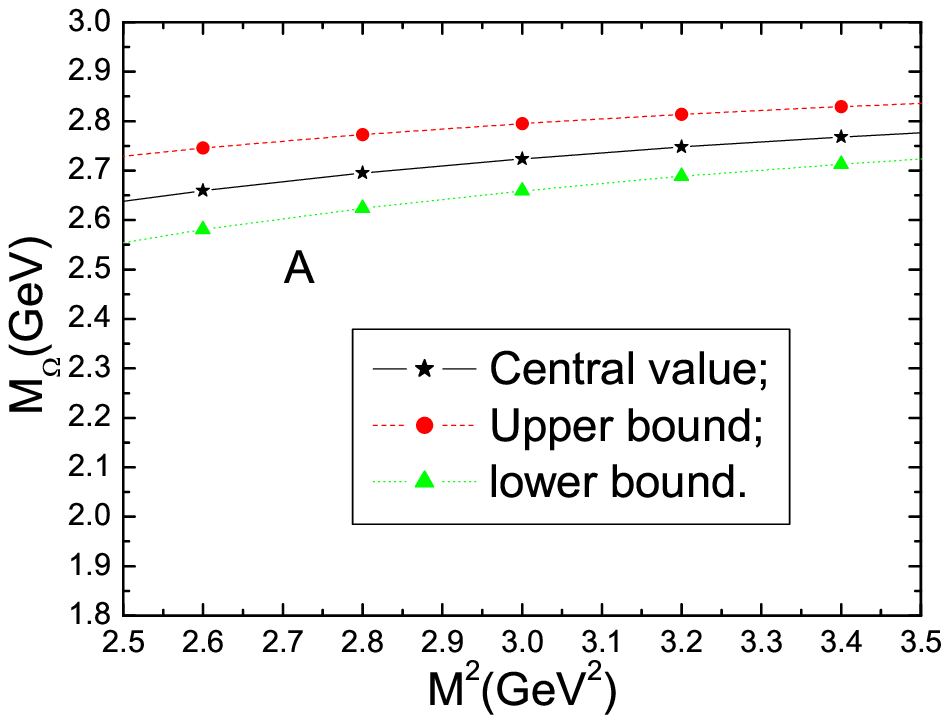}
  \includegraphics[totalheight=6cm,width=7cm]{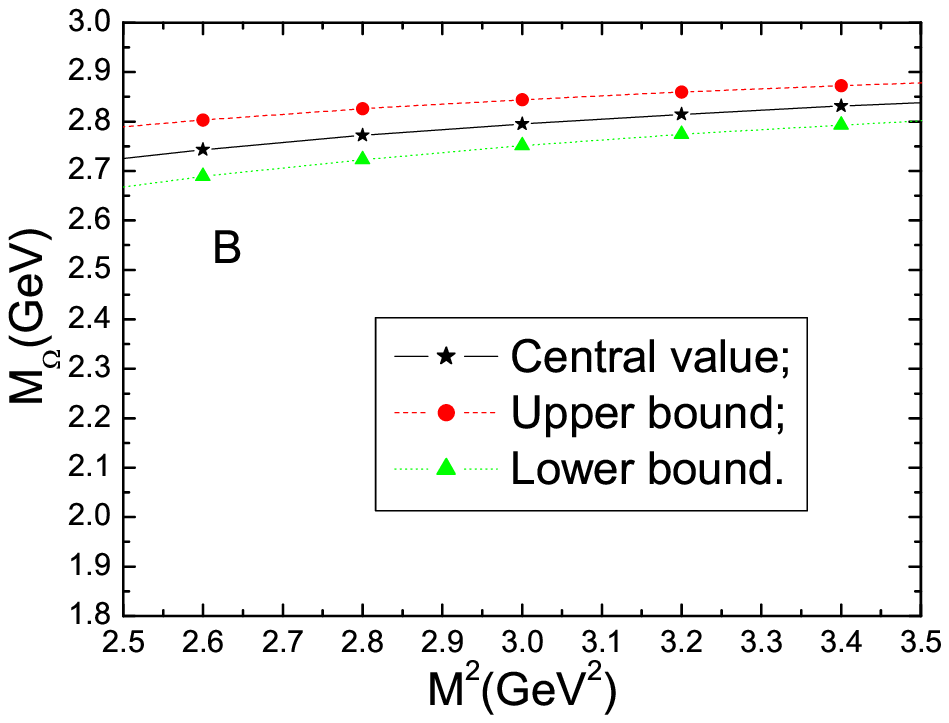}
  \caption{ $M_{\Omega_c^*}$ with  Borel parameter $M^2$, A from Eq.(13) and B from Eq.(14). }
\end{figure}

\begin{figure}
 \centering
 \includegraphics[totalheight=6cm,width=7cm]{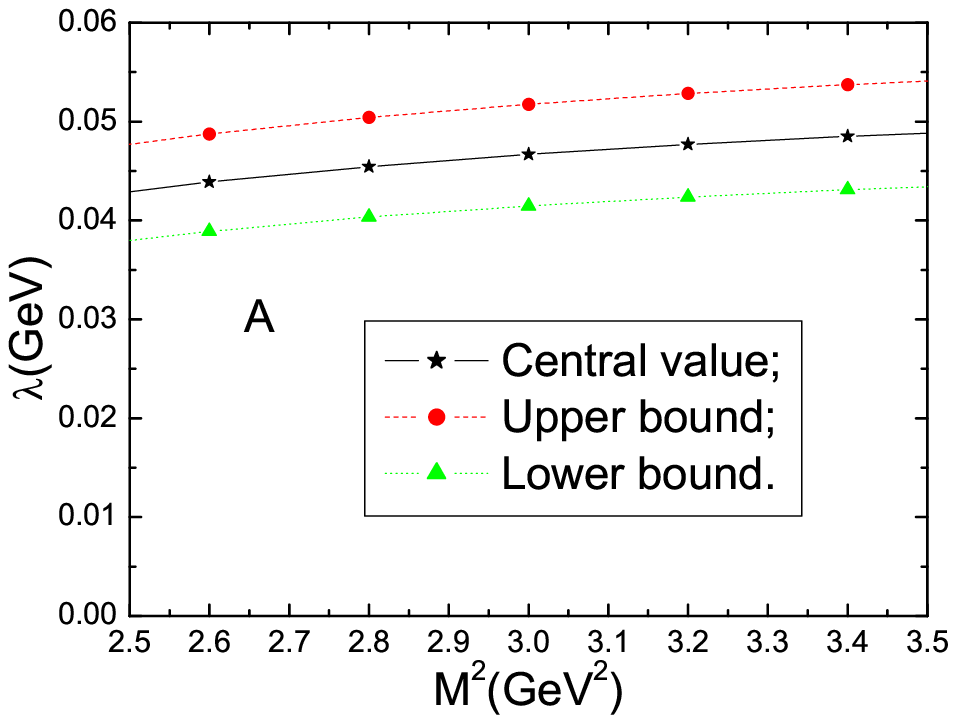}
 \includegraphics[totalheight=6cm,width=7cm]{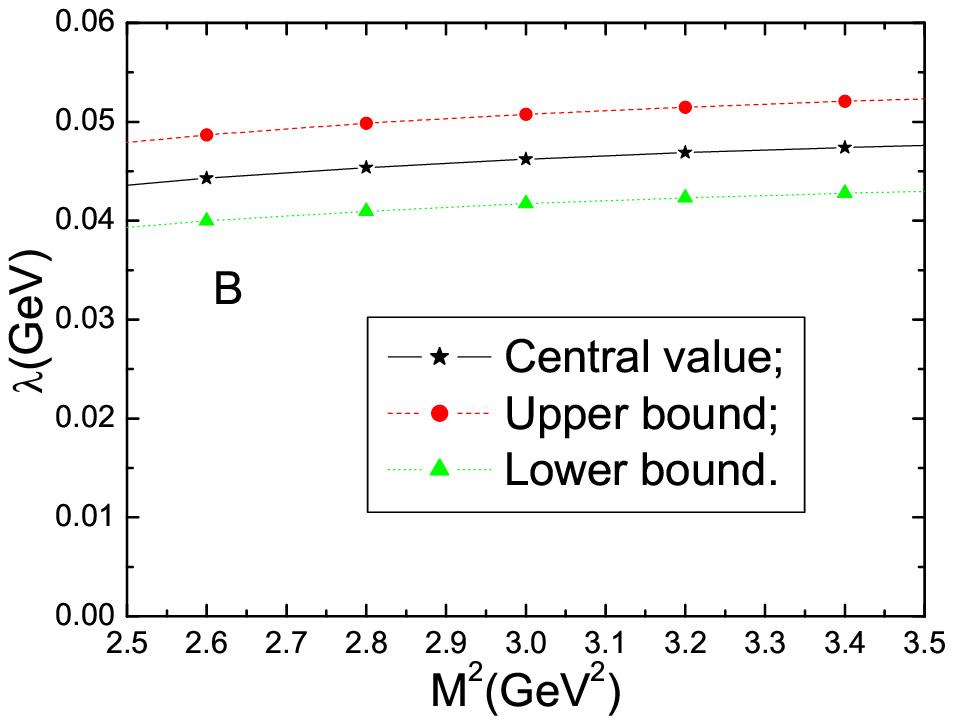}
  \caption{$\lambda_{\Omega_c^*}$ with  Borel parameter $M^2$, A from Eq.(11) and Eq.(13),  and B from Eq.(12) and Eq.(14). }
\end{figure}

\begin{figure}
 \centering
 \includegraphics[totalheight=6cm,width=7cm]{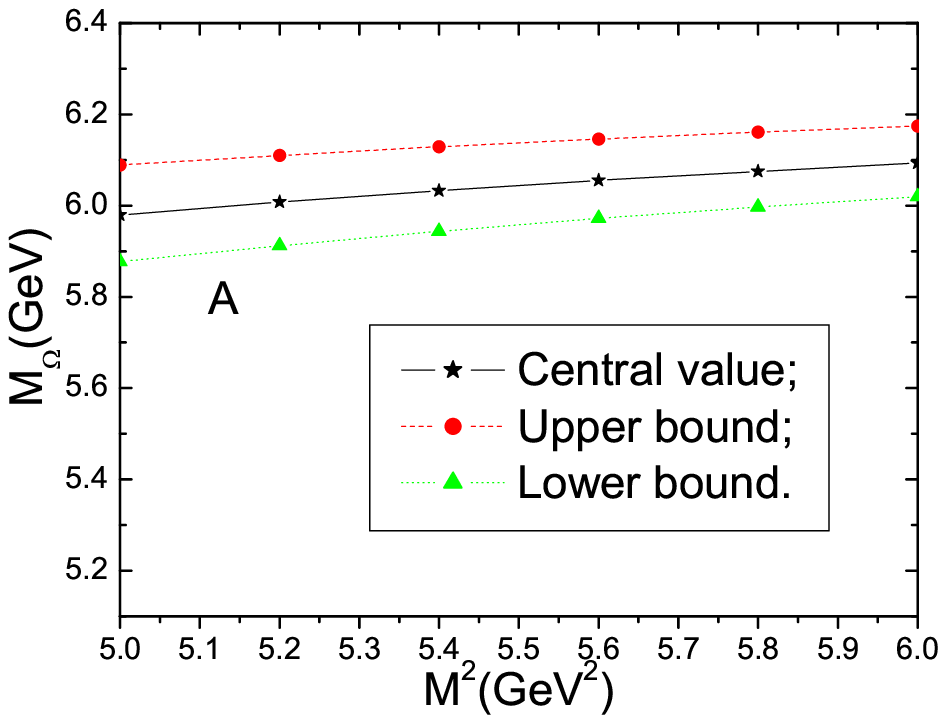}
  \includegraphics[totalheight=6cm,width=7cm]{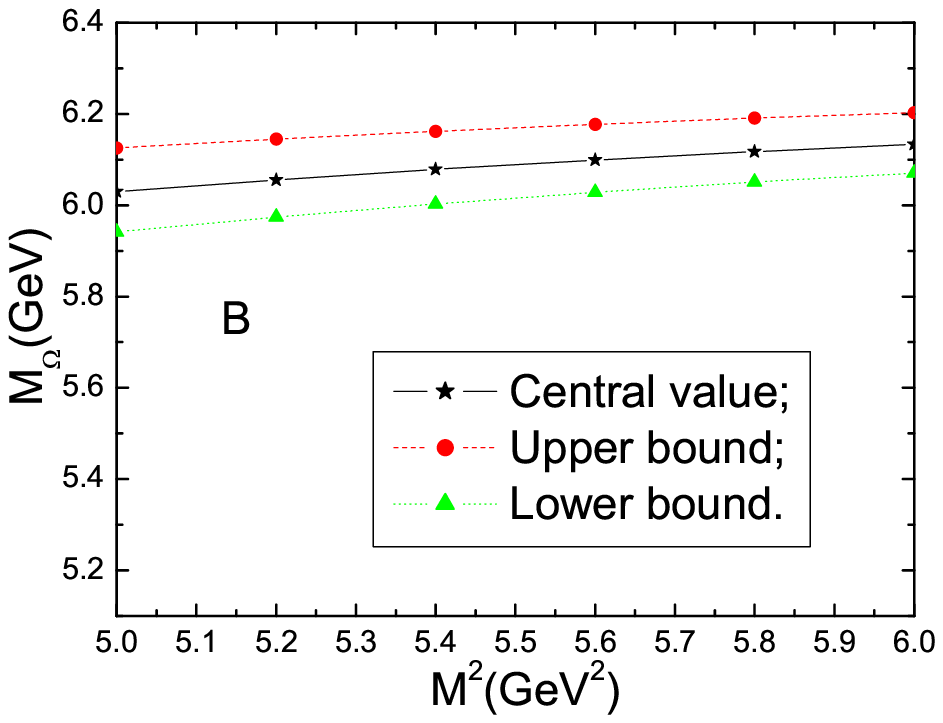}
  \caption{ $M_{\Omega_b^*}$ with  Borel parameter $M^2$, A from Eq.(13) and B from Eq.(14). }
\end{figure}

\begin{figure}
 \centering
 \includegraphics[totalheight=6cm,width=7cm]{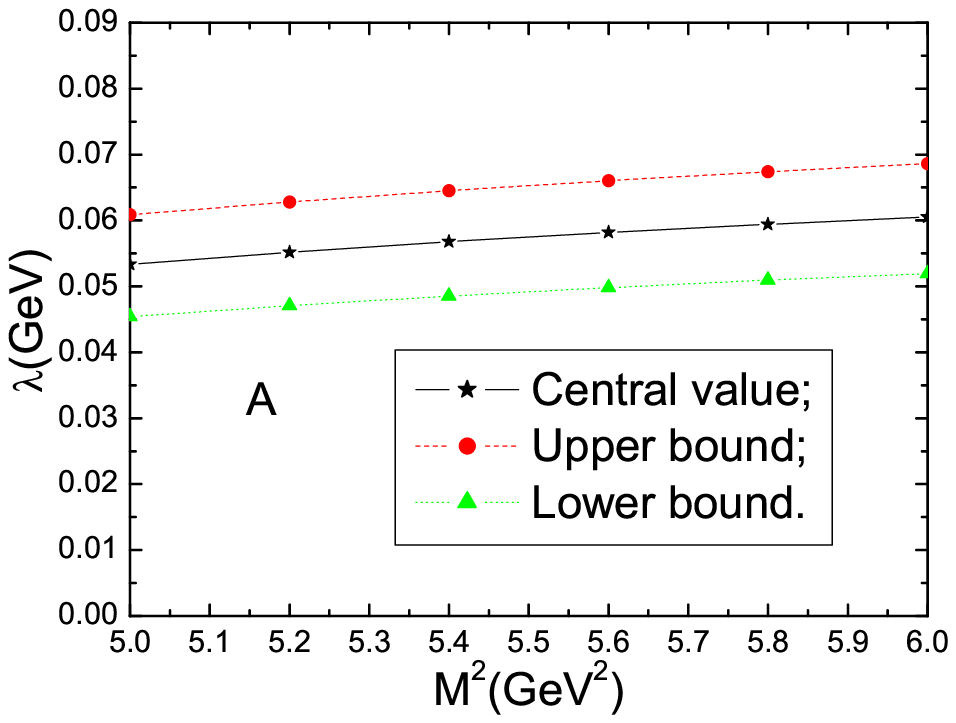}
 \includegraphics[totalheight=6cm,width=7cm]{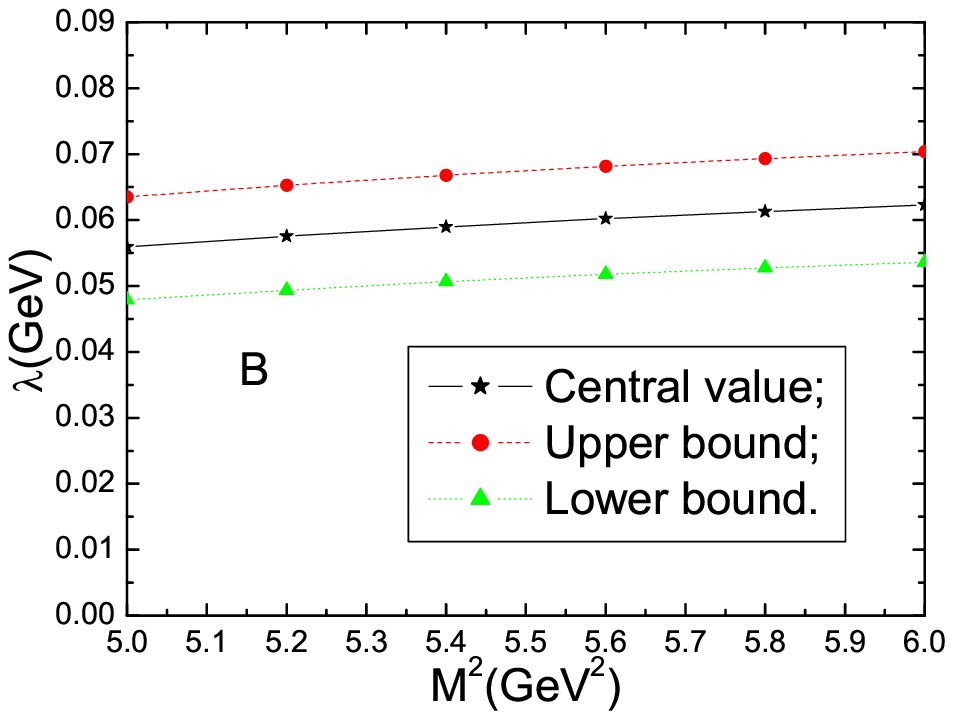}
  \caption{$\lambda_{\Omega_b^*}$ with  Borel parameter $M^2$, A from Eq.(11) and Eq.(13),  and B from Eq.(12) and Eq.(14). }
\end{figure}

Taking into account all uncertainties of the input parameters,
finally we obtain the values of the masses and residues of
 the heavy baryons $\Omega_c^*$ and $\Omega_b^*$, which are
shown in Figs.1-4 respectively,
\begin{eqnarray}
M_{\Omega_c^*}&=&(2.72\pm0.12)\rm{GeV} \, , \nonumber\\
M_{\Omega_b^*}&=&(6.04\pm0.13)\rm{GeV} \, , \nonumber\\
\lambda_{\Omega_c^*}&=&(0.047\pm 0.008)\rm{GeV} \, , \nonumber\\
\lambda_{\Omega_b^*}&=&(0.057\pm 0.011)\rm{GeV} \, ,
\end{eqnarray}
from A from Eq.(11) and Eq.(13), and
\begin{eqnarray}
M_{\Omega_c^*}&=&(2.80\pm0.08)\rm{GeV} \, , \nonumber\\
M_{\Omega_b^*}&=&(6.08\pm0.12)\rm{GeV} \, , \nonumber\\
\lambda_{\Omega_c^*}&=&(0.046\pm 0.007)\rm{GeV}\, , \nonumber\\
\lambda_{\Omega_b^*}&=&(0.060\pm 0.011)\rm{GeV} \, ,
\end{eqnarray}
from   Eq.(12) and Eq.(14). The average values are about
\begin{eqnarray}
M_{\Omega_c^*}&=&(2.76\pm0.10)\rm{GeV} \, , \nonumber\\
M_{\Omega_b^*}&=&(6.06\pm0.13)\rm{GeV} \, , \nonumber\\
\lambda_{\Omega_c^*}&=&(0.047\pm 0.008)\rm{GeV} \, , \nonumber\\
\lambda_{\Omega_b^*}&=&(0.058\pm 0.011)\rm{GeV} \, .
\end{eqnarray}

The value of the mass $M_{\Omega_c^*}$ is compatible with the
experimental data $M_{\Omega_c^*}=(2.768\pm0.003)\rm{GeV}$
\cite{PDG}, the interpolating current $J^{c}_\mu(x)$ can couple with
the charmed baryon $\Omega_c^*$ and give reasonable mass. The value
of the mass $M_{\Omega_b^*}$ for the  bottomed baryon $\Omega_b^*$
with $\frac{3}{2}^+$ is compatible with other theoretical
calculations, $M_{\Omega_b^*}=(6.04-6.09)\rm{GeV}$, such as the
quark models and lattice QCD \cite{QuarkMassB,LattMassB}. Once
reasonable values of the residues $\lambda_{\Omega_c^*}$ and
$\lambda_{\Omega_b^*}$ are obtained, we can take them as   basic
input  parameters and study the hadronic processes \cite{Cheng06},
for example, the radiative decay $\Omega_c^*\rightarrow \Omega_c
\gamma$, with the light-cone QCD sum rules or the QCD sum rules in
external field.

\section{Conclusion}
In this article, we calculate the masses and residues
  of the heavy baryons $\Omega_c^*(css)$ and $\Omega_b^*(bss)$
   with the QCD sum rules. The  numerical values are
  compatible  with the experimental data and other theoretical
  estimations. Once  reasonable values of the residues
$\lambda_{\Omega_c^*}$ and $\lambda_{\Omega_b^*}$ are obtained, we
can take them as  basic parameters and study the  hadronic
processes, for example, the radiative decay $\Omega_c^*\rightarrow
\Omega_c \gamma$, with the light-cone QCD sum rules or the QCD sum
rules in external field.

\section*{Acknowledgments}
This  work is supported by National Natural Science Foundation,
Grant Number 10405009, 10775051, and Program for New Century
Excellent Talents in University, Grant Number NCET-07-0282, and Key
Program Foundation of NCEPU.

\end{document}